\documentclass[]{svmult}

\usepackage{mathptmx}       
\usepackage{helvet}         
\usepackage{courier}        
\usepackage{type1cm}        
\usepackage{makeidx}         
\usepackage{graphicx}        
\usepackage{multicol}        
\usepackage[bottom]{footmisc}

\usepackage{amsmath}
\usepackage{amssymb}
\usepackage{enumerate}

\def\er{Erd\H{o}s-R\'enyi}

\def\eg{e.~g.}

\makeindex             

\begin{document}

\title*{A unified approach to percolation processes on multiplex networks}

\author{G. J. Baxter,  D. Cellai, S. N. Dorogovtsev, A. V. Goltsev and J. F. F. Mendes}

\institute{G.J. Baxter \at Departamento de F\'isica \& I3N, University
  of Aveiro, Campus Universit\'ario de Santiago, 3810-193 Aveiro, Portugal, \email{gjbaxter@ua.pt}
}

\maketitle


\abstract{Many real complex systems cannot be represented by a single
  network, but due to multiple sub-systems and types of interactions,
  must be represented as a multiplex network. This is a set of nodes
  which exist in several layers, with each layer having its own kind
  of edges, represented by different colours. An important fundamental
  structural feature of networks is their resilience to
  damage, the percolation transition. Generalisation of these concepts
  to multiplex networks requires careful definition of what we mean by
  connected clusters. We consider two different definitions. One, a
  rigorous generalisation of the single-layer definition leads to a
  strong non-local rule, and results in a dramatic change in the
  response of the system to damage. The giant component collapses
  discontinuously in a hybrid transition characterised by avalanches
  of diverging mean size. We also consider another definition, which
  imposes weaker conditions on percolation and allows local
  calculation, and also leads to different sized giant components
  depending on whether we consider an activation or pruning
  process. This 'weak' process exhibits both continuous and
  discontinuous transitions. 
}

 
\section{Introduction}

Networks are a powerful tool to represent the
heterogeneous structure of interactions in the study of complex
systems \cite{dgm2008}. But in many cases there are multiple kinds of
interactions, or multiple interacting sub-systems that cannot be adequately
represented by a single network.  Examples include financial
\cite{caccioli2012,Huang2013}, infrastructure \cite{Rinaldi2001}, informatic
\cite{leicht2009} and ecological \cite{Pocock2012} systems.

There are many representations of multi-layer networks, appropriate in
different circumstances(see {\eg} \cite{kivela2013} for a summary). We
focus on multiplex networks, which
are networks with a single set of nodes present in all layers,
connected by a different type of edge (which may be represented by different
colours) in each layer. See \cite{Boccaletti2013} for a recent review
of the topic. Some interdependent networks, in which different layers
have different sets of nodes as well, but the nodes are connected
between layers by interdependency links \cite{Buldyrev2010, Gao2011},
are able to be captured by this construction \cite{Son2012}.

One of the fundamental structural properties of a network is its
response to damage, that is, the percolation transition, where the
giant connected component collapses.
In multi-layer networks, interdependencies between layers can
make a system more fragile. Damage to one element can trigger
avalanches of failures that spread through the whole system
\cite{Osorio2007,Poljansek2012}. Typically a
discontinuous hybrid phase transition is
observed \cite{baxter2012}, similar to those observed in the network
$k$-core or in bootstrap percolation \cite{baxter2011} in contrast to
the continuous transition seen in classical percolation on a simplex
network.
 
Under a weaker definition of percolation, a more complex
phase diagram emerges, with the possibility for both continuous and
discontinuous transitions. When invulnerable or seed nodes are
introduced, we can define activation and pruning processes, which have
different phase diagrams.
The results presented in this Chapter are based on those obtained in
\cite{baxter2012} and \cite{Baxter2014}.

In a single-layer network (simplex), two nodes are connected if there
is at least one path between them along the edges of
the network. A group of connected nodes forms a cluster. The giant
connected component (GCC) is a cluster which contains a finite fraction of
the nodes in the network. The existence of such a giant component is
synonymous with percolation.
We can study it's appearance by applying random damage to the
network. A fraction $\-p$ of nodes are removed, independently at
random, and we check whether the remaining network contains a giant
connected component. Typically the GCC appears linearly with a continuous
second-order transition, although when the
degree distribution is very broad (as in scale-free networks) the
nature and location of the transition may be dramatically altered
\cite{Cohen2002}.

\begin{figure}[t]
\includegraphics[width=0.5\linewidth]{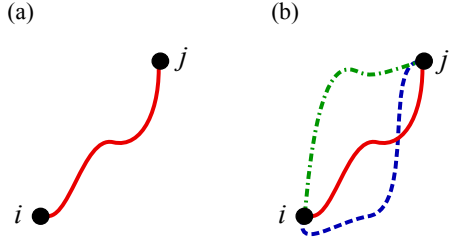}
\caption{(a) In an ordinary network, two vertices $i$ and $j$ belong
  to the same cluster if there is a path connecting them.
(b) In a multiplex network, vertices $i$ and $j$  belong to the same viable cluster if there is a path connecting them for every kind of edge, following
  only edges of that kind. In the example shown, there are $m=3$ kinds
  of edges. Vertices $i$ and $j$ are said to be
  $3$-connected.}\label{m-connected}
\end{figure}

For multiplex networks, we must generalise these definitions of
clusters and percolation.
Consider a multiplex network, with nodes $i = 1, 2, ..., N$
connected by $m$ colours of edges labeled $s = a, b, ..., m $. 
Two nodes $i$ and $j$ are $m$-connected if for each of the $m$
types of edges, there is a path from $i$ to $j$ following edges only
of that type.
Let us suppose that the connections are essential to the
function of each site, so that a node is only viable if it maintains
connections of every type to other viable vertices. A viable cluster
is then a cluster of $m$-connected nodes. This definition is described
in Figure \ref{m-connected}.

In a large system, we wish to find when there is a giant cluster of
viable nodes. From this definition of viable clusters, it follows
that any giant viable cluster is a subgraph of the giant connected
component of each of the $m$ layers formed by considering only a
single colour of edge in the multiplex network. The absence of a giant
connected component in any one of the layers means the absence of the
giant viable cluster. 
Note that when $m=1$, the viable clusters are identical to clusters of
connected vertices in ordinary networks with a single type of edges.
As we will see, the rigorous requirements for viability in multiplex
networks have a profound effect on the percolation of the network,
revealing a discontinuous hybrid transition in the collapse of the
giant viable component.

The viable clusters are related to 
giant mutually connected component in interdependent
networks \cite{Buldyrev2010, Gao2011}. Consider two networks in which a node in
one network may have a mutual dependency on a node in the other
network -- if one is damaged, the other is automatically damaged. To
be part of the giant mutually connected component, a node must be
connected to the cluster via links within its own network, and also
have any interdependency links intact. This system can be mapped to a
multiplex network by simply merging the interdependent nodes into a
single node. Nodes without interdependencies then only have (and
require) links of a single colour \cite{Son2012}. In this way, the
giant viable cluster corresponds to the giant mutually connected
component in the case of full interdependency. When the
interdependency is only partial, the giant mutually connected
component is larger than the giant viable cluster.

If we relax the criterion that a cluster must be connected by all
layers, instead requiring only connection via paths of any colour of
mixture of colours, we immediately return to ordinary percolation,
equivalent to projecting all the layers of the multiplex onto a single
layer, that is, ignoring all the colours. 
If, rather, we were to consider clusters of nodes in which each pair
is connected by at least one single coloured path, the resulting giant
connected component would be the union of the connected components of
the individual layers.

\begin{figure}[t]
\includegraphics[width=0.5\linewidth]{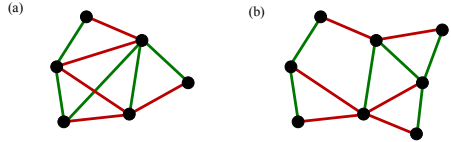}
\caption{Examples of small connected clusters in the strong and weak
  definitions of connectedness in a two-layer multiplex network. (a)
  In the strong definition of a cluster, every node in a viable cluster can
  reach every other by every kind of edge. (b) In the weak
  definition, every node has connections of both colours, but there is
  not necessarily a path of every colour between every pair of
  nodes.}\label{cluster_examples}
\end{figure}

Instead, we may consider a more interesting definition, which is still
weaker than the viable clusters defined above. To differentiate it
from the definition above, we will call this {\emph weak} percolation.
We continue with the requirement that each node only functions if it
is connected to other functioning nodes by every colour of
edge. However, it does not need to be connected to every node in the
cluster by every kind of edge. Weak percolation can be defined in the
following way: a node $i$ is active if, for each of the $m$ colours,
it is connected to at least one active neighbour by an edge of that
colour. Weak percolating clusters are then simply connected clusters
of active nodes. Examples of the connected clusters resulting from the two
different rules can be compared in Fig. \ref{cluster_examples}.

We can consider an activation process, in which a small number of nodes
are initially activated, and activation may spread to neighbouring
nodes. This  can represent, for example, social mobilisation or the
repair of infrastructure after a disaster \cite{Osorio2007}. This
generalizes activation processes such as bootstrap percolation
\cite{bdgm2010} to multiplex networks. Comparing with the
counterpart pruning process, we find that the two processes do not
result in the same giant active component \cite{Baxter2014}.

In the following Section, we analyse the strong definition of
percolation on multiplex networks, identifying the nature of the
percolation transition and the associated avalanches of damage. In
Section \ref{weak}, we analyse the weak definition of percolation and
explore the activation and pruning processes, showing
that they also exhibit hybrid transitions, and outlining the complex
phase diagrams that appear.

\section{Multiplex Percolation}\label{strong}

\begin{figure}[ht]
\includegraphics[width=1.0\linewidth]{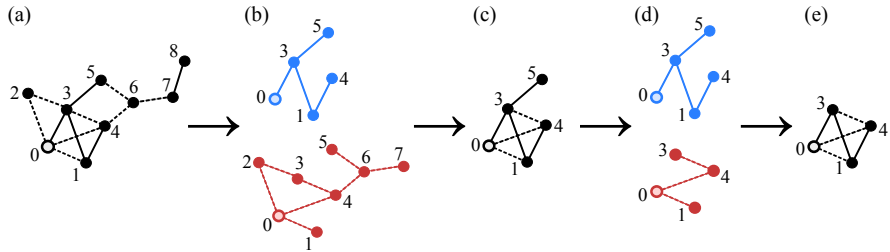}
\caption{An example demonstrating the algorithm for identifying a
  viable cluster in a small network with two kinds of edges. (a) In
  the original network, in step (i) we select vertex $0$ as the test
  vertex. (b) In step (ii) we identify the clusters of vertices
  connected to 0 by each kind of edge. (c) Step (iii): the intersection of these
  two clusters becomes the new candidate set for the viable
  cluster to which 0 belongs. (d)  We repeat steps (ii) using ony
  vertices from the candidate set shown in (c). Repeating step (iii),
  we find the overlap between the two clusters from (d), shown in
  (e). Further repetition of steps (ii) and (iii) does not change this
  cluster, meaning that the cluster consisting of vertices $0$, $1$,
  $3$ and $4$ is a viable cluster. 
}\label{algorithm}
\end{figure}

The viable clusters in a multiplex network can be identified by an
iterative pruning process, testing the connectivity in every layer,
and removing nodes that fail. Such removals may affect the
connectivity of the remaining nodes, so we must repeat the process
until an equilibrium is reached. An algorithm for identifying viable
clusters is the following:

\begin{enumerate}[(i)]
\item Choose a test vertex $i$ at random from the network.
\item For each kind of edge $s$, compile a list of vertices that can be
reached from $i$ by following only edges of type $s$.
\item The intersection of these $m$ lists forms a new candidate set for
the viable cluster containing $i$.
\item Repeat steps (ii) and (iii) but traversing only the current candidate
set. When the candidate set no longer changes, it is either a
viable cluster, or contains only vertex $i$.
\item To find further viable clusters, remove the viable cluster of $i$
from the network (cutting any edges) and repeat steps (i)-(iv) on the
remaining network beginning from a new test vertex.
\end{enumerate}

Note that this process is non-local: it is not possible to identify
whether a node is a member of a viable cluster simply by examining its
immediate neighbours. An example of the use of this algorithm in a
small network is illustrated in Fig. \ref{algorithm}.

We now study in more detail the collapse of the giant viable cluster
under damage by random removal of nodes. We use the fraction $p$ of
undamaged nodes as a control variable. In uncorrelated random
networks the giant viable cluster collapses 
at a critical undamaged fraction $p_c$ in 
a discontinuous hybrid transition, similar
to that seen in the $k$-core or bootstrap percolation
\cite{Dorogovtsev2006a,bdgm2010}.

\begin{figure}[ht]
\includegraphics[width=0.7\linewidth]{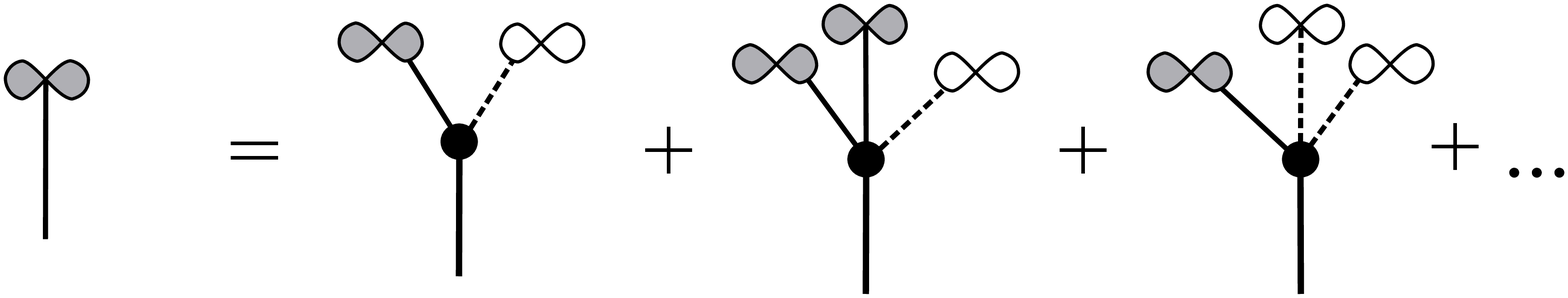}
\caption{Diagrammatic representation of Eq. (\ref{Psi_s}) in a system
  of two interdependent networks $a$ and $b$. The probability $X_a$,
  represented by a shaded infinity symbol
can be written recursively as a sum of second-neighbor
  probabilities. Open infinity symbols represent the equivalent
  probability $X_b$ for network $b$, which obeys a similar recursive
  equation. The filled circle represents the probability $p$ that the
  vertex remains in the network.
}\label{Xa_diagram}
\end{figure}

Let us consider sparse uncorrelated
networks, which are locally tree-like in the infinite size limit $N
\to \infty$.
We take advantage of this locally tree-like property to define
recursive equations which allow us to find the giant viable cluster.
We define $X_s$, with the index $s \in \{a,b,...\}$,
to be the probability that, on following an
arbitrarily chosen edge of type $s$, we encounter the root of an
infinite sub-tree formed solely from type $s$ edges, whose vertices
are also each connected to at least one infinite subtree of every
other type. We call this a type $s$ infinite subtree. The vector
$\{X_a,X_b,...\}$ plays the role of the order parameter.
In a two-layer network, for example, the probability $X_a$ can be
written as the sum of second-level probabilities in terms of $X_a$ and
$X_b$, as illustrated in Figure \ref{Xa_diagram}. 
In general, writing this graphical representation in equation form, using
the joint degree distribution $P(q_a,q_b,...)$, we arrive at the self
consistency equations (for more details, see \cite{baxter2012})
\begin{align}\label{Psi_s}
X_s =&
p\sum_{q_a,q_b,...}\!\!\!\!\frac{q_s}{\langle q_s\rangle}
P(q_a,q_b,...) \big[1-(1-X_s)^{q_s-1} \big] \!\prod_{l\neq s}\!
\big[1-(1-X_l)^{q_l}\big]\nonumber\\
 \equiv & \Psi_s(X_a,X_b,...)\,,
\end{align}
where $p$ is the probability that the vertex was not initially damaged.
The term $({q_s}/{\langle q_s\rangle}) P(q_a,q_b,...)$ gives the
probability that on following an arbitrary edge of type $s$, we
find a vertex with degrees $q_a, q_b,...$, while
$[1-(1\!-\!X_a)^{q_a}]$ is the probability that this vertex has at
least one edge of type $a \neq s$ leading to the root of an
infinite sub-tree of type $a$ edges. This
becomes $[1-(1\!-\!X_s)^{q_s-1}]$ when $a = s$.

\begin{figure}[ht]
\includegraphics[width=0.7\linewidth]{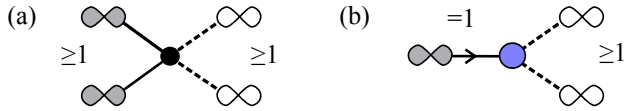}
\caption{Viable and critical viable vertices for two interdependent
  networks.
(a) A vertex is in the giant viable cluster if it
  has connections  of both kinds to giant viable subtrees, represented
  by infinity symbols, which occur with probabilities $X_a$ (shaded)
  or $X_b$ (open) -- see text.
(b) A critical viable vertex of type $a$ has exactly $1$
  connection to a giant sub-tree of type $a$.
}\label{active_and_critical}
\end{figure}

A vertex is then in the giant viable cluster if it has at least
one edge of every type $s$ leading to an infinite type $s$ sub-tree
(probability $X_s$), as shown in Fig.~\ref{active_and_critical}(a)
\begin{equation}\label{S}
S = p\sum_{q_a,q_b,...}\!\! P(q_a,q_b,...)\!\prod_{s=a,b,...}\!\!
\big[1-(1\!-\!X_s)^{q_s} \big],
\end{equation}
which is equal to the relative size of the giant viable cluster of the
damaged network.

A hybrid transition appears at the point where $\Psi_s(X_a,X_b,...)$ first
meets $X_s$ at a non-zero value, for all $s$.
This occurs when
\begin{equation}\label{hybrid_condition_asym}
\det[{\bf J}-{\bf I}] = 0
\end{equation}
 where ${\bf I}$ is the unit matrix and ${\bf J}$ is the Jacobian matrix
$J_{ab} = \partial \Psi_b/\partial X_a$
. 
The critical point $p_c$ can then be found by simultaneously solving
Eqs. (\ref{Psi_s}) and (\ref{hybrid_condition_asym}).
To find the scaling near the critical point, we expand
Eq. (\ref{Psi_s}) about the critical value $X_s^{(c)}$. We find that
\begin{equation}\label{squareroot}
X_s - X_s^{(c)} \propto (p-p_c)^{1/2}.
\end{equation}
This square-root scaling is the typical behaviour of the order
parameter near a hybrid transition. It results from avalanches of
spreading damage which diverge in size near the transition.
The scaling of the size of the giant viable cluster, $S$,
immediately follows
\begin{equation}
S - S_c \propto (p-p_c)^{1/2}.
\end{equation}

\subsection{Avalanches}

We now examine the avalanches of damage which occur in the system, 
 in order to understand the nature of the transition more completely.
We focus on the case of two types of edges.
Consider a viable node that has exactly one edge of type $a$ leading
to a type $a$ infinite subtree, and at least one edge of type $b$
leading to a type $b$ infinite subtree. We call this a
critical node of type $a$. It is illustrated in
Fig.~\ref{active_and_critical}(b). It is a critical vertex because it
will be removed from the viable cluster if it loses its 
 single link to a type $a$ infinite subtree. The removal of any node
 from the giant viable cluster, and the edges to which it is
 connected, therefore also requires the removal of  any critical
 vertices which depend on the removed edges. Removed critical nodes
 may have edges leading to further critical nodes. This is the way
 that damage propagates in the system. The removal of a single node
 can result in an avalanche of removals of critical vertices from the
 giant viable cluster.

\begin{figure}[ht]
\includegraphics[width=0.5\linewidth]{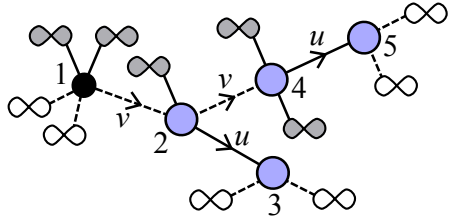}
\caption{
A critical cluster. Removal of any of the shown viable vertices will
result in the removal of all downstream critical viable
vertices. Vertices $2$-$5$ are critical vertices.
Removal of the vertex labeled $1$
will result in all of the shown vertices being removed (becoming
non-viable). Removal of vertex $2$ results in the removal of vertices
$3$, $4$, and $5$ as well, while removal of vertex $4$ results only in
vertex $5$ also being removed. As before, infinity symbols represent
connections to infinite viable subtrees. Other connections to
non-viable vertices or finite viable clusters are not shown.
}\label{critical_cluster}
\end{figure}

To represent this process visually, we draw a diagram of viable
nodes and the edges between them.
We mark the special critical edges, that critical viable nodes depend on,
with an arrow leading to the critical node. An avalanche can only
transmit in the direction of the arrows. For example, in
Fig.~\ref{critical_cluster}, removal of the vertex labeled 1 removes
the essential edge of the critical vertex 2 which thus becomes
non-viable. Removal of vertex 2 causes the removal of further critical
vertices 3 and 4, and the removal of 4 then requires the removal of 5.
Thus critical vertices form critical clusters. 
Graphically, upon removal of a vertex,
we remove all vertices found by following the arrowed
edges, which constitutes an avalanche.
Note that an avalanche is a branching process. Removing a
  vertex may lead to avalanches along several edges emanating from the
  vertex (for example, in Fig. \ref{critical_cluster}, removing vertex
  2 leads to avalanches along two edges). 
As we approach the critical point from above, the avalanches
increase in size. 
The mean size of avalanches triggered by a randomly removed vertex
finally diverges in size at the critical point,
which is the cause of the discontinuity in the size of the giant
viable cluster, which collapses to zero. These avalanches are thus an
inherent part of a hybrid transition. To show this, we use a
generating function approach \cite{Newman2001} to calculate the sizes
and structure of avalanches.

\begin{figure}[ht]
\includegraphics[width=0.50\linewidth]{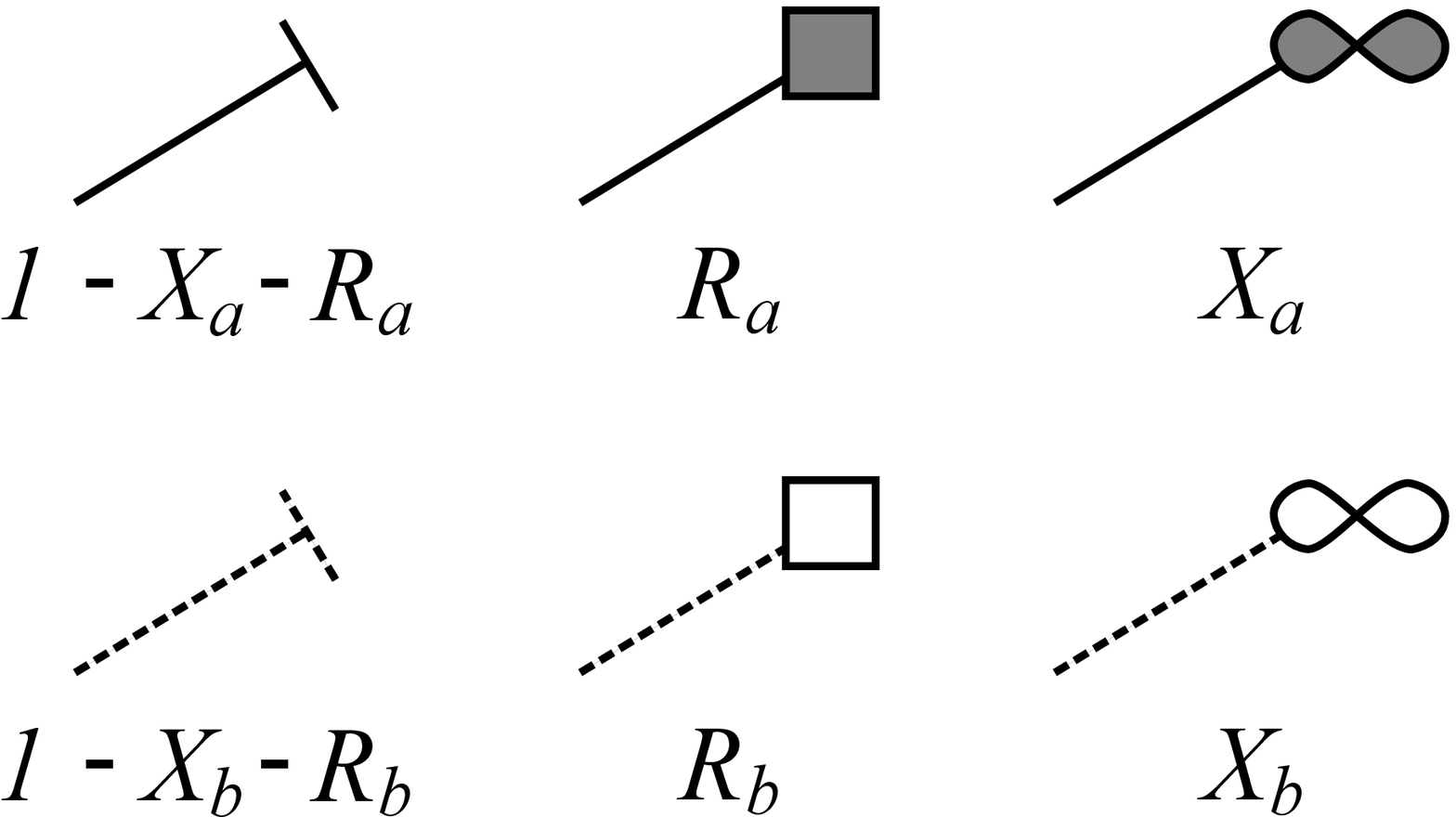}
\caption{Symbols used in the diagrams to represent key
  probabilities. Solid lines represent edges of type $a$, dashed lines
represent edges of type $b$.}\label{symbols}
\end{figure}

There are three possibilities when following an
arbitrarily chosen edge of a given type: i) with probability $X_s$ we
encounter a type $s$ infinite subtree  ii) with probability
$R_s$ we encounter a vertex which has a connection to an infinite
subtree of the opposite type, but none of the same type. Such a vertex
is part of the giant viable cluster if the parent vertex was; or
iii) with probability $1-X_s-R_s$, we encounter a vertex which has no
connections to infinite subtrees of either kind.
These probabilities are represented graphically in
Fig. \ref{symbols}. We will use these symbols in subsequent diagrams.

\begin{figure}[ht]
\includegraphics[width=0.8\linewidth]{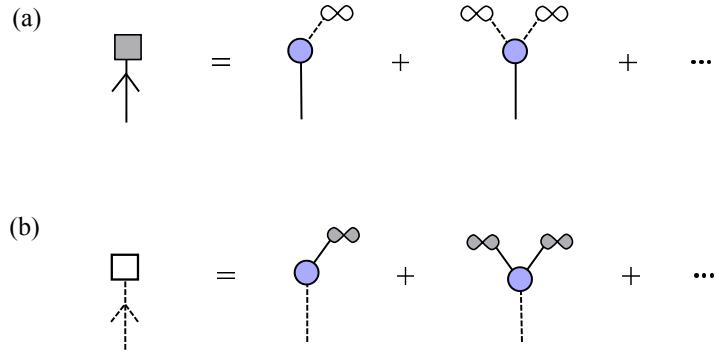}
\caption{(a) The probability $R_a$ can be defined in terms of the
  second-level connections of the vertex found upon following an edge
  of type $a$. Note that possible connections to `dead ends' -- vertices not in
  the viable cluster (probability $1-X_a-R_a$ or $1-X_b - R_b$) are
  not shown.  (b) The equivalent graphical equation for the
  probability $R_b$.}\label{R_diagram}
\end{figure}

The probability $R_a$ obeys
\begin{equation}
R_a = \sum_{q_a}\sum_{q_b}\! \frac{q_a}{\langle q_a\rangle}P(q_a,q_b)
(1\!-\!X_a)^{q_a-1} \left[1\! -\! (1\!-\!X_b)^{q_b}\right]
\label{R1}
\end{equation}
 and similarly for $R_b$. This equation is represented graphically in
 Fig. \ref{R_diagram}.

\begin{figure}
\includegraphics[width=0.95\linewidth]{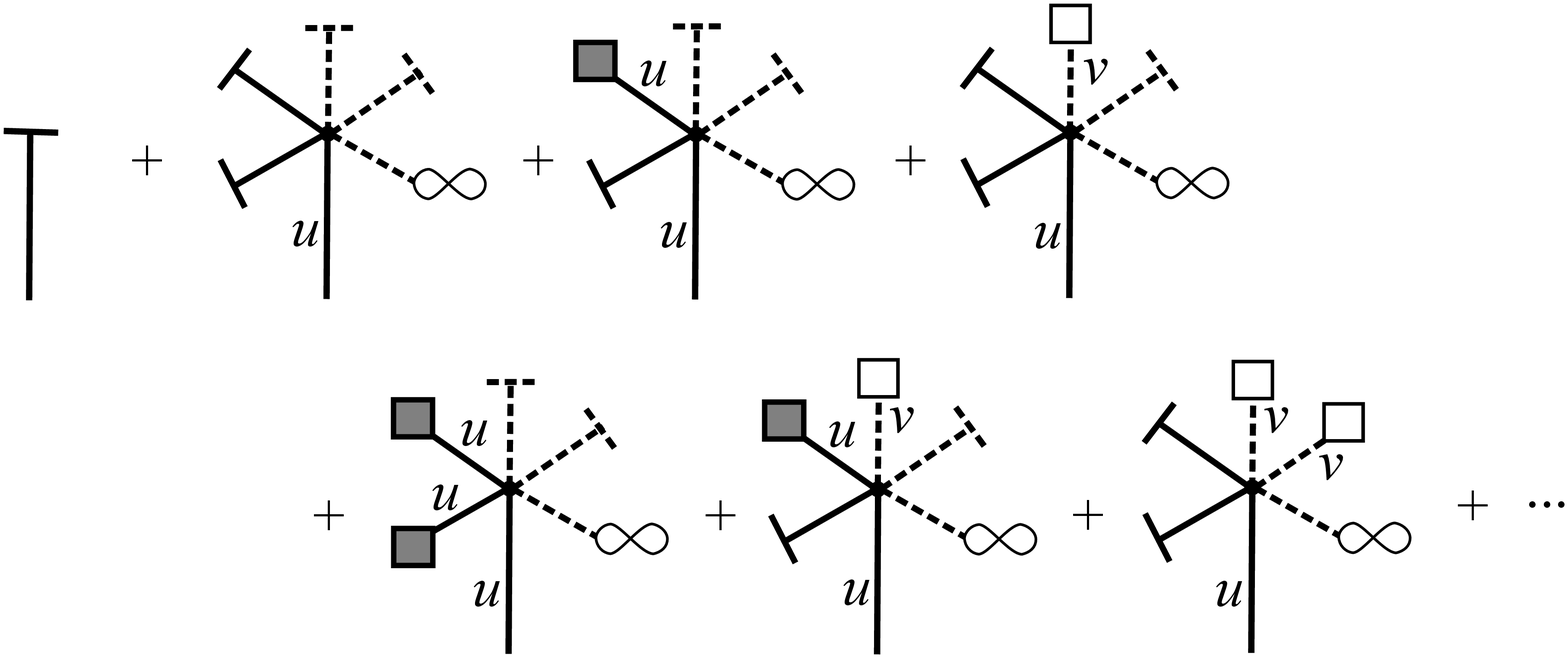}
\caption{Representation of the generating function $H_{a}(x,y)$ (right-hand side of Eq. \ref{Ha_asym}) for the
  size of a critical cluster encountered upon following an edge of
  type $a$.}\label{H_asym_diagram}
\end{figure}

The generating function for the size of the avalanche triggered by removing
an arbitrary type $a$ edge which does not lead to an infinite type $a$
subtree can be found by considering the terms represented in Figure
\ref{H_asym_diagram}.
The first term represents the probability that, upon following an edge of
type $a$ (solid lines) we reach a node
with no connection to a type $b$ subtree (and hence is not
viable),that is, a critical cluster of size $0$. The second
term represents the probability to encounter a critical cluster of
size $1$. The node encountered
has a connection to the type $b$ infinite subtree (infinity symbol),
but no further connections to viable nodes. Subsequent terms
represent recursive probabilities that the vertex encountered has $1$
(third and fourth terms), $2$ (fifth, sixth, seventh terms) or more
connections to further potential critical clusters.
The variables $u$ (for type $a$ edges) and $v$ (type $b$) are assigned
to each such edges.

Considering these terms, we can write the generating function for
the number of critical vertices encountered upon following an
arbitrary edge of type $a$ (that is, the size of the
resulting avalanche if this edge is removed) as
\begin{equation}\label{Ha_asym}
H_{a}(u,v) = 1-X_a-R_a + uF_a[H_{a}(u,v),H_{b}(u,v)]
\end{equation}
and similarly for $H_{b}(u,v)$, the corresponding generating function
for the size of the avalanche caused by removing a type $b$ edge is
\begin{equation}\label{Hb_asym}
H_{b}(u,v) = 1-X_b-R_b + vF_b[H_{a}(u,v),H_{b}(u,v)],
\end{equation}
where the
functions $F_a(x,y)$ and $F_b(x,y)$ are defined as:
\begin{align}
F_{a}(x,y) \equiv & \sum_{q_a}\sum_{q_b}\! \frac{q_a}{\langle
  q_a\rangle}P(q_a,q_b) x^{q_a-1}\sum_{r=1}^{q_b}\!\binom{q_b}{r} X_b^r
y^{q_b-r}
\label{F1_asym}
\end{align}
and similarly for $F_{b}(x,y)$, by exchanging all subscripts $a$ and
$b$.
While the function $F_a(x,y)$ does not necessarily represent a
physical quantity or probability, we can see that it incorporates the
probability of encountering a vertex with at least one child edge of
type $b$ leading to a giant viable subtree (probability $X_b$) upon
following an edge of type $a$. All other outgoing edges then
contribute a factor $x$ (for type $a$ edges) or $y$ (type $b$).
Here $u$ and $v$ are auxiliary variables. Following through a critical
cluster, a factor $u$ appears for each arrowed edge of type $a$, and
$v$ for each arrowed edge of type $b$. For example, the critical
cluster illustrated in Fig.~\ref{critical_cluster} contributes a
factor $u^2 v^2$.

The mean number of critical vertices reached upon
following an edge of type $a$, i.e. the mean size of the resulting
avalanche if this edge is removed, is given by
$\partial_u H_a(1,1)+\partial_v H_a(1,1)$, where $\partial_u$
signifies the partial derivative with respect to $u$.
Unbounded avalanches emerge at the point where $\partial_u H_a(1,1)$
[or $\partial_v H_{b}(1,1)$] diverges.
Taking derivatives of Eq. (\ref{Ha_asym}),
\begin{align}
\partial_u H_{a}(u,v) =&
F_{a}[H_{a},H_{b}] + u\left\{\partial_u H_{a}
\partial_x F_{a}[H_{a},H_{b}]
+ \partial_u H_{b} \partial_y F_{a}[H_{a},H_{b}] \right\}\\[5pt]
\partial_v H_{a}(u,v) =&
 u\left\{\partial_v H_{a} \partial_x F_{a}[H_{a},H_{b}]
+ \partial_v H_{b} \partial_y F_{a}[H_{a},H_{b}] \right\}
\end{align}
with similar equations for $\partial_u H_{b}(u,v)$
and $\partial_v H_{b}(u,v)$.
Some rearranging gives
\begin{equation}\label{del_H_int1}
\partial_u H_{a}(1,1) =
\frac{R_a + \partial_u H_{b}(1,1)
  \partial_y F_{a}(1-X_a,1-X_b)}
{1 - \partial_x F_{a}(1-X_a,1-X_b)}
\end{equation}
and
\begin{equation}\label{del_H_int2}
\partial_v H_{a}(1,1) =
\frac{\partial_u H_{a}(1,1)
\partial_x F_{b}(1-X_a,1-X_b)}
{1 - \partial_y F_{b}(1-X_a,1-X_b)}
\end{equation}
where we have used that $H_{a}(1,1) = 1-X_a$ and $F_{a}(1-X_a,1-X_b)
= R_a$.
From Eqs. (\ref{Psi_s}) and (\ref{F1_asym}),
\begin{align}
\partial_x F_{a}(1-X_a,1-X_b) =&
\frac{\partial}{\partial X_a}\Psi_a(X_a,X_b)\\
\partial_y F_{b}(1-X_a,1-X_b) =&
\frac{\langle q_a\rangle}{\langle q_b\rangle} \frac{\partial}{\partial
  X_a}\Psi_b(X_a,X_b) ,
\end{align}
and similarly for $\partial_x F_{b}$ and
$\partial_y F_{b}$,
which when substituted into (\ref{del_H_int1}) and
(\ref{del_H_int2}) give
\begin{equation}\label{del_Ha}
\partial_u H_{a}(1,1) =
\frac{  R_a[1- \frac{\partial}{\partial X_b} \Psi_{b}(X_a,X_b)]}
{\det[{\bf J}-{\bf I}]}\,.
\end{equation}
We see that the denominator
exactly matches the left-hand side of
Eq. (\ref{hybrid_condition_asym}), meaning that
the mean size of avalanches triggered by random removal of vertices diverges
exactly at the point of the hybrid transition.

\section{Weak Multiplex Percolation}\label{weak}

Now we consider, for comparison, the weaker definition of percolation
on multiplex networks. In this case we also find a discontinuous
hybrid transition, but a continuous second order transition may also
occur.

In ordinary percolation, and the strong
multiplex percolation considered above, 
activation and deactivation yield the same
giant cluster. In weak percolation, however, activation of the
network yields a very different phase diagram than a pruning
process. We define an activation process, which we call Weak
Bootstrap Percolation (WBP) and a deactivation/pruning process, Weak
Pruning Percolation (WPP). We also introduce invulnerable vertices,
which are always active. These are necessary to seed the activation
process, and we include them in the pruning process, for symmetry.

\subsection{Weak Pruning Percolation (WPP)}

Let us begin with Weak Pruning Percolation.
A fraction $f$ of the nodes are randomly assigned as invulnerable, the
rest being vulnerable. In
the WPP process, the network is then damaged, with a fraction $p$ of
all nodes remaining undamaged. Once again, $p$ acts as a control
parameter. Each of the remaining vulnerable nodes is pruned if it
fails to have at least one connection in each layer to a surviving
node (vulnerable or invulnerable). The removal of some nodes may
affect the neighbourhoods of other surviving nodes, so the pruning
process must be repeated until no more nodes can be
removed. Invulnerable nodes cannot be pruned.

Let $Z_s$ be the probability that, upon following an edge of type $s$,
we encounter the root of a sub-tree (whether finite or infinite)
formed solely from type $s$ edges, whose vertices are also each
connected to at least one such subtree of every other type.
We define $X_s$ as the probability that such a subtree is
infinite. Precisely,  $X_s$ is the probability that each member the
subtree encountered, as well as meeting the criteria for $Z_s$, also
has at least one edge leading to an infinite subtree of any type
(probability $X_a$ etc.).

In a multiplex with $m$ types of edges and a degree distribution
$P(q_a,q_b,...)$, the equation for $Z_s$ is (see \cite{Baxter2014} for
more details):
\begin{equation}
	Z_{s} = p f + p (1-f) \sum_{q_a,q_b,...} \frac{q_s
          P(q_a,q_b,...)}{\langle q_s\rangle}\prod_{n\neq s}
\left[1 - (1-Z_n)^{q_n} \right] \equiv \Phi_s(Z_a,Z_b,...).
	\label{eq:Z-wpp-general}
\end{equation}
The first term ($pf$) accounts for the probability that the
encountered node is an undamaged invulnerable node, which is always
active, and so its state doesn't depend on any of its neighbours.
The second term (proportional to $p(1-f)$) calculates the recursive
probability for vulnerable undamaged nodes.

The equation for $X_s$ is
\begin{align}
\begin{split}
    X_{s} =& p f  \sum_{q_a,q_b,...} \frac{q_s P(q_a,q_b,...)}{\langle
      q_s\rangle} \big[ 1 -(1-X_s)^{q_s-1} \prod_{n\neq s}
    (1-X_n)^{q_n} \big] \\
 &+ p(1-f)\!\!\!\!\! \sum_{q_a,q_b,...}\!\!\!\!\!\frac{q_s P(q_a,q_b,...)}{\langle q_s\rangle}
\Bigg\{
\prod_{n\neq s}
    [1{-}(1-Z_n)^{q_n}] 
 {-}  (1\!-\!X_s)^{q_s\!-1}\\
& \times \prod_{n\neq s} [(1-X_n)^{q_n}{-}(1-Z_n)^{q_n}] \Bigg\}
\end{split}\nonumber\\
\equiv& \Psi_s(X_a,X_b,...,Z_a,Z_b,...)&.
    \label{eq:X-wpp-general}
\end{align}
The first sum on the right hand side calculates the probability that
we encounter an undamaged invulnerable node, and which has at least
one child edge leading to an infinite subtree of any type.
The second sum calculates the same probability but in the case when the encountered node is not invulnerable.
This term is written as a difference between the probability of having
at least one edge leading to finite or infinite subtrees of each type
and another term which removes the possibility that all of the
subtrees are finite.
This last product must be multiplied by $(1-X_s)^{q_s-1} $ to
exclude the possibility of reaching an infinite subtree by a type $s$ edge.

Finally, having given equations for $Z_s$ and $X_s$, we can use them
to find $S$, the probability that a randomly chosen
node is in the giant percolating cluster defined in this model. This
is the strength of the giant percolating cluster.
It is given by the following formula:
\begin{multline}
	S = p f  \sum_{q_a,q_b,...} P(q_a,q_b,...) \left[ 1 -  \prod_s
	(1-X_s)^{q_s} \right]\\[5pt] 
	   + p (1-f) \sum_{q_a,q_b,...} P(q_a,q_b,...) 
           \prod_s \left[ 
	1 - (1-Z_s)^{q_s} \right] - \prod_s  \left[ 
	  (1-X_s)^{q_s} - (1-Z_s)^{q_s} \right].
	\label{eq:S-wpp-general}
\end{multline}
This equation is constructed in a similar way to that for $X_s$.

A continuous transition appears at the point where a non-zero solution
to $X_s = \Psi_s$ first appears.
A hybrid transition appears at the point where $\Psi_s$
is first tangent to $X_s$ at a non-zero value, for all $s$. Because a
jump in $X_s$ is always accompanied by a jump in $Z_s$, it is more
simple to look for the point where $\Phi_s$ is tangent to $Z_s$.
This occurs when
\begin{equation}\label{hybrid_condition_wpp}
\det[{\bf J}-{\bf I}] = 0
\end{equation}
where ${\bf I}$ is the unit matrix and ${\bf J}$ is the Jacobian matrix
$J_{ab} = \partial \Phi_b/\partial X_a$.
Together these criteria allow us to map the phase diagram of the
process with respect to the two parameters $f$ and $p$.

\begin{figure}[htb]
\begin{center}
	\includegraphics[width=0.5\columnwidth]{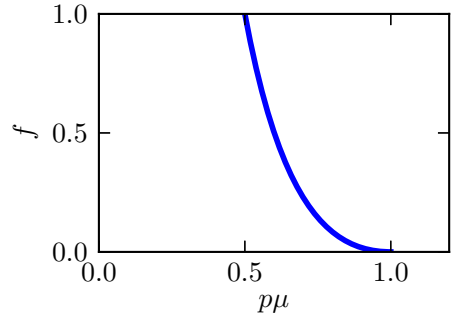}
	\caption{Phase diagram of the WPP model for two uncorrelated {\er} networks with identical mean degree $\mu$.}
	\label{fig:wpp-ph-diag}
\end{center}
\end{figure}

In the case of only two layers, the phase diagram is characterized
by a line of continuous phase transitions. An example is shown in
Fig.~\ref{fig:wpp-ph-diag}, for the case where each of the two layers
is an {\er} network, with identical mean degree $\mu$.
In the limit $f=0$, the probability of a node being in the giant WPP component is given by the product of the classical percolation probability in each layer.
In the {\er} example shown in the figure, this means the percolation
point is at $\nu \equiv p\mu = 1$.
In the limit $f=1$, all nodes are invulnerable, and the situation
corresponds to classical percolation with the multiplex is treated as
a single network.
There is no hybrid transition in the two layer case.

\begin{figure}[htb]
\begin{center}
	\includegraphics[width=0.7\columnwidth,angle=0]{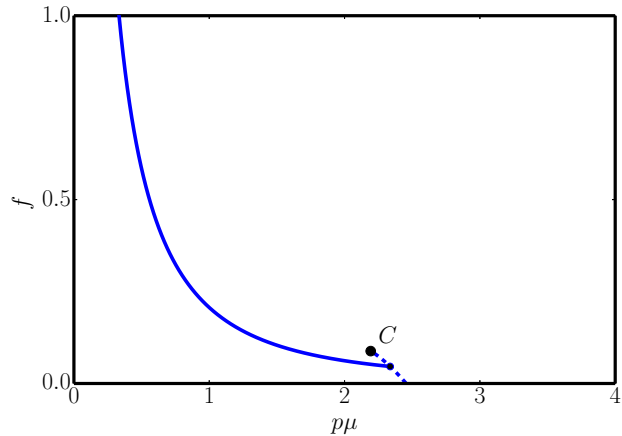}
	\caption{Phase diagram of the WPP model for three uncorrelated
          {\er} networks with identical mean degree $\mu$. The solid
          line gives the location of the continuous transition, the
          dashed line gives the location of the discontinuous
          transition. The point $C$ is the critical point.}
	\label{fig:wpp-3er-ph-diag}
\end{center}
\end{figure}

In the case of three layers, now a hybrid transition also appears. 
The line of discontinuous transitions can be calculated by solving
Eqs. (\ref{eq:Z-wpp-general}) and (\ref{hybrid_condition_wpp})
together.
An example phase diagram is given in
Fig.~\ref{fig:wpp-3er-ph-diag}. We see that the both continuous and
discontinuous transitions are present, with the giant component
appearing discontinuously for small $f$, and having two transitions
for slightly larger $f$: a continuous appearance followed by a
discontinuous hybrid transition.

\subsection{Weak Bootstrap Percolation (WBP)}

Now we consider an activation process called Weak Bootstrap Percolation, which extends the concept of bootstrap percolation \cite{chalupa1979} to
multiplex networks. As for the pruning model, a fraction $f$ of nodes
are invulnerable, and are active from the start. Again, a random
damage is applied to the network, with the undamaged fraction $p$
acting as the control parameter. Now, however, the vulnerable nodes
begin in an inactive state. A node becomes active if it has at least
one active neighbour in each of the $m$ layers of the multiplex. The
activation of nodes may in turn provide the required active neighbours
to more nodes, so the process is repeated until no more nodes can
become active. 

\begin{figure}[htb]
	\includegraphics[width=0.5\columnwidth]{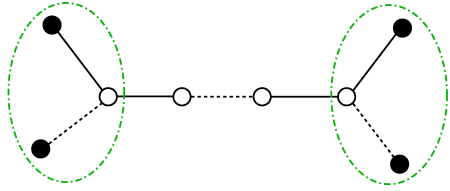}
	\caption{Example of clusters in a multiplex with two types of edges.
	Black nodes are invulnerable/seed vertices, white nodes are vulnerable vertices.
	In WPP, all the nodes are unprunable (remain active), because each white node
        is connected to another node by each edge type.
	In WBP, only the nodes inside the green dot-dashed lines become active, while the remaining two nodes have only one active neighbor, by one edge type only, so they cannot become active.}
	\label{fig:difference-wpp-wbp}
\end{figure}

At the end of the activation process, the active clusters are in
general not the same as those that would be found through the pruning
process. This is because in WPP nodes are considered active until
pruned. This means that, for example, a pair of nodes connected by an edge
of one type, provide the required support of that type for one
another, even if neither has another edge of that type. In WBP, on the
other hand, such an isolated dimer can never become activated
(Fig.~\ref{fig:difference-wpp-wbp}). The same holds for many larger
configurations as well.

Let $Z_s$ be the probability that, upon following an edge of type $s$,
we encounter the root of a sub-tree (whether finite or infinite)
formed solely from type $s$ edges, whose vertices are also each
connected to at least one such subtree of every type. This obeys the
equation
\begin{align}
	Z_{s} &= p f + p (1-f) \sum_{q_a,q_b,...} \frac{q_s
          P(q_a,q_b,...)}{\langle q_s\rangle} \left[ 1 -
          (1-Z_s)^{q_s-1}\right]  
         \prod_{n\neq s}
	\left[1 - (1-Z_n)^{q_n} \right]\nonumber\\
&\equiv \Phi_s(Z_a,Z_b,...).
	\label{eq:Z-wbp-general}
\end{align}
This differs from the equivalent equation for WPP,
(\ref{eq:Z-wpp-general}) because now each node must have connections
of every type, not just of the types different from $s$.

Similarly, we define $X_s$ as the probability that such a subtree is
infinite. Precisely,  $X_s$ is the probability that each member the
subtree encountered, as well as meeting the criteria for $Z_s$, also
has at least one edge leading to an infinite subtree of any type.

An argument similar to the one for Eq.~(\ref{eq:X-wpp-general}) leads
us to the equation:
\begin{multline}
	X_{s} = p f  \sum_{q_a,q_b,...} \frac{q_s P(q_a,q_b,...)}{\langle q_s\rangle} \bigg[ 1 - (1-X_s)^{q_s-1} \prod_{n\neq s}
	(1-X_n)^{q_n} \bigg] \\
+ p(1-f) \sum_{q_a,q_b,...}\frac{q_s P(q_a,q_b,...)}{\langle q_s\rangle}
\bigg\{ [ 1 - (1-Z_s)^{q_s-1}] \prod_{n\neq s}
	[1-(1-Z_n)^{q_n}] \\
 - [(1-X_s)^{q_s-1} - (1-Z_s)^{q_s-1}] 
\prod_{n\neq s}
	[(1-X_n)^{q_n}-(1-Z_n)^{q_n}] \bigg\}.
	\label{eq:X-wbp-general}
\end{multline}

\begin{figure}[htb]
\begin{center}
	\includegraphics[width=0.8\columnwidth]{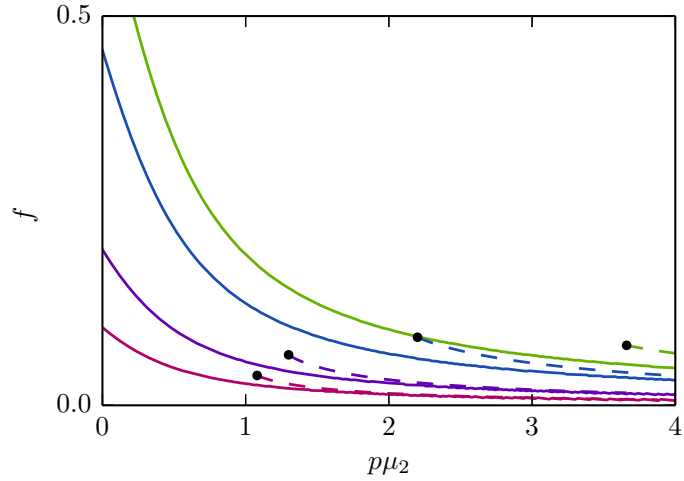}
	\caption{Phase diagram of the WBP model for two uncorrelated
          {\er} networks with mean degree $\mu_1$ and
          $\mu_2$. Horizontal axis is $\nu_2 = p\mu_2$. Each solid
          curve shows the location of the continuous transition for a
          particular value of $\nu_1$, from top to bottom $\nu_1 =
          \{1.5, 2.193, 5, 10\}$.
          Dashed curves show the corresponding location of the
          discontinuous transition (which is always above the
          continuous transition), with circles marking the critical
          end point. Color online.}
	\label{fig:wbp-ph-diag_ERasym}
\end{center}
\end{figure}

While $Z_n$ and $X_n$ are different from their WPP counterparts, the
equation for $S$ is the same as Eq. (\ref{eq:S-wpp-general}).
In the case of WBP, a hybrid transition appears already in a two layer
multiplex. A typical phase diagram is plotted in Figure
\ref{fig:wbp-ph-diag_ERasym}, for the case of two {\er}
layers with different mean degrees. Now we see that the giant
component always first appears
continuously, with a second, discontinuous hybrid transition occurring
afterwards, for small $f$. The line of discontinuous transitions is
found using the conditions
\begin{equation}
	\left\{%
		\begin{array}{lcr}
			\Phi_{f,\nu_1,\nu_2}(z) &=& 1\\
			\Phi^{\prime}_{f,\nu_1,\nu_2}(z) &=& 0
		\end{array}
	\right.
	\label{eq:wbp-asym-disc}
\end{equation}
The line ends at the critical point defined by these two conditions in
combination with a third condition
\begin{equation}
\Phi^{\prime\prime}_{f,\nu_1,\nu_2}(z) = 0.
\end{equation}

\subsection{Avalanches}
\label{sec:clusters}

To understand the discontinuous transitions which we observe in the
two weak percolation models, we again analyze avalanches, which
propagate through clusters of critical vertices.
 Diverging avalanche sizes lead to the discontinuous transitions.
As before, in the pruning process, WPP, a critical vertex is a vertex that only just meets the criteria for
inclusion in the percolating cluster (in the case of WPP). However, in
the activation process, WBP, the avalanches which diverge in mean size
at the discontinuous transition are of activations of nodes,
not of pruning, so critical nodes are those that just fail
to meet the criteria for activation. 

\begin{figure}[htb]
\begin{center}
	\includegraphics[width=0.6\columnwidth]{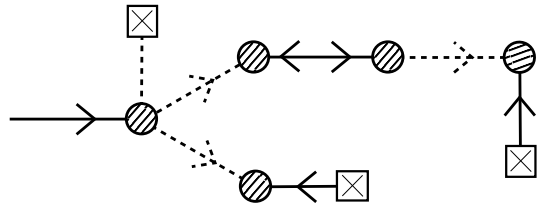}
	\caption{A representation of a cluster of critical vertices in
          WPP. Hatching indicates that vertices are members of the WPP
          percolating cluster. Because critical vertices are in the percolating
          cluster for WPP, a critical vertex may be linked to the
          percolating cluster via another critical vertex. That is,
          external edges of type $Z_s$ are not necessarily
          required. Furthermore, this means that critical dependencies
          can be bi-directional: it is possible for avalanches to
          propagate in either direction along such edges. Note that
          outgoing critical edges must be of the opposite type to the
          incoming one. The boxes containing crosses represent the
          probability $Z_n-R_n$.}
	\label{fig:wpp-avalanche}
\end{center}
\end{figure}

In the case of WPP, a
critical node of type $s$ has exactly one connection to an infinite
subtree of type $s$, and at least one of all the other types. A vertex
may be critical with respect to more than one type, if it
simultaneously has exactly one connection to infinite subtrees of different
types.
Such a vertex is related to avalanches because it has one (or
possibly more) edge(s) which, if lost, will cause the vertex to be
pruned from the cluster. If, in turn, other outgoing edges of this vertex 
are critical edges for other critical vertices, these vertices
will also be removed. Chains of such connections therefore delineate
the paths of avalanches of spreading damage. An example is shown in
Fig. \ref{fig:wpp-avalanche}. Damage to the node at one end of an
edge is transmitted along arrowed edges.

There are three possibilities when following an
arbitrarily chosen edge of a given type: i) with probability $X_s$ we
encounter a type $s$ infinite subtree  ii) with probability
$R_s$ we encounter a vertex which has a connection to an infinite
subtree of the opposite type, but none of the same type. Such a vertex
is part of the giant viable cluster if the parent vertex was; or
iii) with probability $1-X_s-R_s$, we encounter a vertex which has no
connections to infinite subtrees of either kind.
These probabilities are represented graphically in
Fig. \ref{symbols}. We will use these symbols in subsequent diagrams.

To examine these avalanches, we define the probability $R_s$, to be
the probability that, on following an edge of type $s$, we encounter a
vulnerable vertex (probability $1-f$), which has not been
removed due to random damage (probability $p$) and has at least one child
edge of each type $n \neq s$ leading to a subtree defined by the
probability $Z_n$, and zero of type $s$. That is
\begin{equation}\label{eq:R1_WPP}
R_s = p(1-f) \sum_{q_a,q_b,...}\frac{q_s P(q_a,q_b,...)}{\langle q_s\rangle}
(1-Z_s)^{q_s-1} \prod_{n\neq s}
\left[ 1 - (1-Z_n)^{q_n}\right].
\end{equation}
 We can then define a generating function for the size of the critical
subtree encountered upon following an edge of type $s$ (and hence
resulting pruning avalanche should the parent vertex of that edge
be removed) in a recursive way by
\begin{equation}\label{eq:H1_WPP}
H_s(\vec{u}) = Z_s - R_s + u_s F_s[H_1(\vec{u}), H_2(\vec{u}),...,H_m(\vec{u})].
\end{equation}
Where the functions $F_s(\vec{x})$ are defined to be
\begin{multline}\label{eq:F1_WPP}
F_s(\vec{x}) = p(1-f) \sum_{q_a,q_b,...}\frac{q_s P(q_a,q_b,...)}{\langle q_s\rangle}
(1 - Z_s)^{q_m-1} \prod_{n\neq s}
\sum_{l=1}^{q_s} \binom{q_s}{l} (1 - Z_n)^{q_n-l} x_n^{l}.
\end{multline}
Notice that $F_s$ has no dependence on $x_s$.
This method is very similar to that used in \cite{baxter2012}.
A factor $u_s$ appears for every critical edge of type $s$ appearing
in the subtree. The first terms $Z_s-R_s$ give the probability that
zero critical nodes are encountered. The second term, with factor
$u_s$, counts the cases where the first node encountered is a critical
one. This node may have outgoing edges leading to further critical
nodes. These edges are counted by the function $F_s$, and the use of
the generating functions $H_n$ as arguments recursively counts the
size of the critical subtree reached upon following each of these edges.

The mean size of the avalanche caused by the removal of single vertex
is then given by 
\begin{equation}
\sum_{s} \partial_{u_s}H_s(\vec{1})\,.
\end{equation}
Where $\partial_z$ signifies the partial derivative with respect to
variable $z$.

Let us first examine the mean avalanche size in the case of two layers.
Taking partial derivatives of Eqs. (\ref{eq:H1_WPP}) and
(\ref{eq:F1_WPP}), and after some rearranging, we arrive at
\begin{equation}
\partial_{u_1} H_1(1,1) = \frac{R_1} {1 -
  \partial_{x_2} F_1(Z_1,Z_2)\partial_{x_1}F_2(Z_1,Z_2) }.
\end{equation}
where we have used that $F_1(Z_1,Z_2) = R_1$ and also that
$H_1(1,1) = Z_1$, and $H_2(1,1) = Z_2$.

Let us define the right-hand side of Eq.~(\ref{eq:Z-wpp-general}) to be
$\Psi_1(Z_1,Z_2)$. 
From Eq.~(\ref{eq:Z-wpp-general}), and comparing with Eq. (\ref{eq:F1_WPP}), the
partial derivatives of $\Psi_1(Z_1,Z_2)$, are
\begin{eqnarray}
\frac{\partial\Psi_1}{\partial Z_1} &=& 0 \nonumber\\
\frac{\partial\Psi_1}{\partial Z_2} &=&  p(1-f)\sum_{q_1,q_2}\frac{
  P_{q_1,q_2}}{\langle q_1\rangle} q_1q_2(1-Z_2)^{q_2-1} \nonumber\\
&& =
\frac{\langle q_2\rangle}{\langle q_1\rangle}\frac{\partial}{\partial_{x_1}}F_2(Z_1,Z_2).
\end{eqnarray}
and similarly for ${\partial\Psi_2}/{\partial Z_1}$ and
${\partial\Psi_2}/{\partial Z_2}$.
Substituting back, we find that
\begin{equation}
\partial_u H_1(1,1) = \frac{R_1}{(\partial \Psi_1/\partial
  Z_2)(\partial \Psi_2/\partial
  Z_1)}.
\end{equation}
The denominator remains finite, and the numerator does not diverge, so
this quantity remains finite everywhere in the 2-layer WPP model. This
confirms that a discontinuous transition does not occur when there are
only two layers.

Following the same procedure in the case of three layers reveals that
\begin{equation}
\partial_{u_1} H_1(1,1) = 
R_1 \left\{1 - \frac{\partial_2\Psi_1[\partial_1\Psi_2 +
    \partial_1\Psi_3\partial_3\Psi_2]}
{1 - \partial_2\Psi_3\partial_3\Psi_2} - \partial_1\Psi_3\partial_3\Psi_1
\right\}^{-1}
 = \frac{R_1}{1 - \frac{d \Psi_1}{d Z_1}}.
\end{equation}
where for compactness we have written $\partial_m\Psi_n$ for $\partial
\Psi_n /\partial Z_m$.
Now, an alternative form for the condition for the location of the
discontinuous transition is $\frac{d \Psi_1}{d Z_1} = 1$. We see
immediately that this implies that the mean avalanche size diverges at
the critical point.
 In other words the
avalanches diverge in size as the discontinuous hybrid transition
approaches, just as the susceptibility does for an ordinary
second-order transition. 

\begin{figure}[htb]
\begin{center}
	\includegraphics[width=0.6\columnwidth]{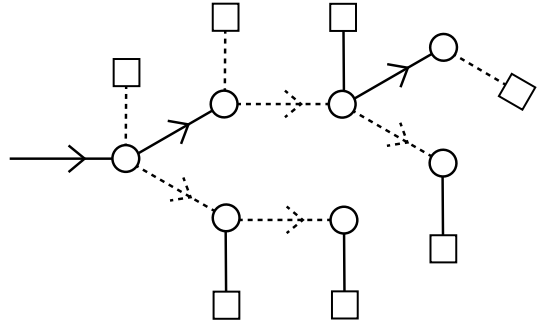}
	\caption{An example of a critical cluster in WBP. Avalanches
          of activation propagate through the cluster following the
          arrowed edges. If an upstream vertex is activated, all
          downstream critical vertices will in turn be activated. Note
        that, unlike for WPP,  in WBP it is not possible for an edge to be arrowed in both directions. Activation can only ever propagate in one
        direction along a given edge. Also note
          that in the WBP case outgoing critical
          edges must be of the same type as the incoming one.}
	\label{fig:wbp-avalanche}
\end{center}
\end{figure}

In the case of the activation process, WBP, a
critical vertex is one that fails the activation criterion for a single type of
edge. That is, it has exactly zero edges leading to the root of type
$s$ subtrees (probability $Z_s$), and at least one of every other type.
If such a node gains a single
connection to the root of a type $s$ subtree, it will itself become the root of
such a subtree. Chains of such connections therefore delineate
the paths of avalanches of spreading activation. An example of a small
critical cluster is shown in Figure \ref{fig:wbp-avalanche}.

To examine these avalanches, we now define the probability $R_s$, to be
the probability that, on following an edge of type $s$, we encounter a
vertex which is not a seed vertex (probability $1-f$), has not been
removed due to random damage (probability $p$) has at least one child
edge of all other types $n\neq s$ leading to the appropriate subtrees (probability $Z_n$), and zero of type $s$. That is
\begin{equation}\label{eq:R1_WBP}
R_s = p(1-f) \sum_{q_a,q_b,...}\frac{q_s P(q_a,q_b,...)}{\langle q_s\rangle}
(1-Z_s)^{q_s-1}  \prod_{
			n\neq s}
\left[ 1 - (1-Z_n)^{q_n}\right]
\end{equation}
Note that this is identical to
(\ref{eq:R1_WPP}), but the probabilities $Z_n$ are different, as is
the following argument.

Because critical vertices are outside the WBP cluster, the
probabilities $Z_s$ and $R_s$ are mutually
exclusive. This means that, upon following an edge of type $s$, there
are three mutuall exclusive possibilities: i) we encounter a subtree
of type $s$ (probability $Z_m$)ii) we encounter a critical vertex
(probability $R_s$)or iii) we encounter neither (probability $1-Z_s-R_s$).
We can then define a generating function for the size of the critical
subtree encountered upon following an edge of type $s$ (and hence
resulting activation avalanche should the parent vertex of that edge
be activated) in a recursive way by
\begin{equation}\label{eq:H1_WBP}
H_s(\vec{u}) = 1 - Z_s - R_s + u_s F_s[H_1(\vec{u}), H_2(\vec{u}),...,H_m(\vec{u})]\,.
\end{equation}
The functions $F_s(\vec{x})$ are defined to be
\begin{equation}\label{eq:F1_WBP}
F_s(x,y) = p(1-f) \sum_{q_a,q_b,...}\frac{q_s P(q_a,q_b,...)}{\langle q_s\rangle}
x_s^{q_s-1} \prod_{
			n\neq s}
 \sum_{l=1}^{q_n} \binom{q_n}{l} Z_n^l x_n^{q_n-l}.
\end{equation}
Note that $F_s(1-Z_1,1-Z_2,...,1-Z_m) = R_s$ and $H_s(\vec{1}) = 1-Z_s$.

The mean size of the avalanche caused by the activation of a single
vertex is again given by 
\begin{equation}
\sum_{s} \partial_{u_s}H_s(\vec{1}).
\end{equation}

Let us consider the case of WBP in a 2-layer multiplex.
Taking partial derivatives of (\ref{eq:H1_WBP}) and (\ref{eq:F1_WBP}) and
after some rearranging, we find
\begin{equation}
\partial_{u_1} H_1(1,1) = 
\frac{R_1\left[1 - \partial_{x_2}F_2
    \right]} {\left[1\! -\! \partial_{x_1}F_1
    \right]\left[1\! -\! \partial_{x_2}F_2
    \right] - \partial_{x_2}F_1
  \partial_{x_1}F_2 }
\end{equation}
where for brevity we have not written the arguments of the derivatives
of the functions $F_1$ and $F_2$, but they should be taken to be evaluated at
$(1-Z_1,1-Z_2)$, and
%
%
where we have used that $F_1(1\!-\!Z_1,1\!-\!Z_2) = R_1$ and also that
$H_1(1,1) = 1- Z_1$, and $H_2(1,1) = 1 - Z_2$.

Remembering that we have defined $\Phi_s(Z_1,Z_2)$, in
the two layer case, to be the right-hand side of Eq.~(\ref{eq:Z-wbp-general}),
\begin{align}
\frac{\partial \Phi_1}{\partial Z_1} &= 
p(1-f)
\sum_{q_1,q_2}\frac{ P_{q_1,q_2}}{\langle q_1\rangle}q_1(q_1-1)
(1-Z_1)^{q_1-2} [1 - (1-Z_2)^{q_2}]\nonumber\\
 &= \partial_{x_1} F_1(1-Z_1,1-Z_2)
\end{align}
and
\begin{align}
\frac{\partial \Phi_1}{\partial Z_2} &= 
p(1-f) 
\sum_{q_1,q_2}\frac{ P_{q_1,q_2}}{\langle q_1\rangle}q_1q_2
(1-Z_2)^{q_2-1} [1 - (1-Z_1)^{q_1-1}]\nonumber\\
&= \frac{\langle q_2\rangle}{\langle q_1\rangle} \partial_{x_1} F_2(1-Z_1,1-Z_2)
\end{align}
and a similar procedure is followed for $\Phi_2$.
This means that the equation for $\partial_{u_1}H_1(1,1)$ can be written
\begin{equation}
\partial_{u_1}H_1(1,1) = \frac{R_1 [1 - \partial \Phi_2 /\partial Z_2]}
        {\det[{\bf J}-{\bf I}]}\,.
\end{equation}
where the Jacobian matrix ${\bf J}$ has elements $J_{ij} = \partial
\Phi_i/Z_j$, and ${\bf I}$ is the identity matrix.
The condition $\frac{d \Phi_1}{d Z_1} = 1$ for the location of the
discontinuity in $Z_1$ (and $Z_2$) can be rewritten
\begin{equation}
\det[{\bf J}-{\bf I}] = 0
\end{equation}
meaning that $\partial_{u_1}H_1(1,1)$ diverges, and hence the mean
avalanche size, diverges precisely at the critical point. This
indicates that indeed a discontinuous hybrid transition, with
accompanying avalanches of activations, appears even in the two-layer
multiplex.
A similar analysis can be performed for three or more layers.

\section{Conclusions}

In conclusion, the study percolation in multiplex networks requires
new definitions of connectivity. 
We have studied the robustness of multiplex networks to
damage under two different definitions of connectedness. In the first,
a natural generalization of the concept of single network
connectedness, we find a strong criterion which leads to an abrupt
collapse of the giant component of a multiplex
network having two or more layers. In contrast to ordinary networks,
where two vertices are
connected if there is a path between them, in multiplex network with $m$
types of edges, two vertices are $m$-connected if for every kind of edge
there is a path from one to another vertex.
The transition is a discontinuous hybrid transition, similar
to that found, for example, in the network $k$-core problem. The
collapse occurs through avalanches which diverge in size when the
transition is approached from above. We described critical clusters
associated with these avalanches. The avalanches are responsible for
both the critical scaling and the discontinuity observed in the size
of the giant viable cluster.

We compared this with a weaker definition of connectedness, but one which
can be calculated locally. In this definition, nodes are members of a
cluster if they have at least one edge of each type leading to another
member of the cluster. This means that two nodes can belong to the
same cluster even when there are no paths of a single colour
connecting them. We also introduced the concept of invulnerable
nodes. In the pruning process form of this model, we find that a
two-layer multiplex network no longer exhibits a hybrid transition in
the collapse of the giant component, but in three layers such a
transition can occur. Finally, we introduced an activation process on
multiplex network, dual to the weak pruning process, in which a small
number of seed (invulnerable) nodes are initially activate and further
nodes activate if they have connections by every type of edge to
active neighbours. The two processes have related phase diagrams, but
we find that a discontinuous hybrid transition can occur even when
there are only two layers.


This work was partially supported by FET IP Project
MULTIPLEX 317532 and by the FCT projects EXPL/FIS-NAN/1275/2013
and PEst-C/CTM/LA0025/2011, and post-doctoral fellowship
SFRH/BPD/74040/2010.

\bibliography{multiplex_chapter}
\end{document}